# Countering Trusting Trust through Diverse Double-Compiling

David A. Wheeler
*Institute for Defense Analyses*
*dwheeler @ ida.org*

## Abstract

*An Air Force evaluation of Multics, and Ken Thompson's famous Turing award lecture "Reflections on Trusting Trust," showed that compilers can be subverted to insert malicious Trojan horses into critical software, including themselves. If this attack goes undetected, even complete analysis of a system's source code will not find the malicious code that is running, and methods for detecting this particular attack are not widely known. This paper describes a practical technique, termed diverse double-compiling (DDC), that detects this attack and some compiler defects as well. Simply recompile the source code twice: once with a second (trusted) compiler, and again using the result of the first compilation. If the result is bit-for-bit identical with the untrusted binary, then the source code accurately represents the binary. This technique has been mentioned informally, but its issues and ramifications have not been identified or discussed in a peer-reviewed work, nor has a public demonstration been made. This paper describes the technique, justifies it, describes how to overcome practical challenges, and demonstrates it.*

## 1. Introduction

Many product security evaluations examine source code, under the assumption that the source code accurately represents the product being examined. Naïve developers presume that this can be assured simply by recompiling the source code to see if the same binary results.

Unfortunately, if an attacker can modify the binary file of the compiler, this is insufficient. An attacker who can control the compiler binary (directly or indirectly) can render source code evaluations worthless, because the compiler can re-insert malicious code into anything it compiles—including itself.

Karger and Schell provided the first public description of the problem. They noted in their examination of Multics vulnerabilities that a "penetrator could insert a trap door into the... compiler... [and] since the PL/I compiler is itself written in PL/I, the trap door can maintain itself, even when the compiler is recompiled. Compiler trap doors are significantly more complex than the other trap doors... However, they are quite practical to implement" [1].

Ken Thompson widely publicized this problem in his famous 1984 Turing Award presentation "Reflections on Trusting Trust," clearly explaining it and demonstrating that this was both a practical and dangerous attack. He first described how to modify the Unix C compiler to inject a Trojan horse, in this case to modify the operating system login program to surreptitiously give him root access. He then showed how to modify and recompile the compiler itself with an additional Trojan horse devised to detect itself. Once this is done, the attacks can be removed from the source code so that no source code—even of the compiler—will reveal the existence of the Trojan horse, yet the attacks could persist through recompilations and cross-compilations of the compiler. He then stated that "No amount of source-level verification or scrutiny will protect you from using untrusted code... I could have picked on any program-handling program such as an assembler, a loader, or even hardware microcode. As the level of program gets lower, these defects will be harder and harder to detect" [2]. As a demonstration, Thompson implemented this attack in the C compiler and successfully subverted another Bell Labs group; the attack was never detected. Thompson's demonstration subverted the login program (for control) and the disassembler (to hide the attack from disassembly). The malicious compiler was never released outside Bell Labs [3].

For source code security evaluations to be strongly credible, there needs to be a way to justify that the source code being examined accurately represents the files being executed—yet this attack subverts that very claim. Internet Security System's David Maynor argues that the risk of these kinds of attacks is increasing [4][5]; Karger and Schell noted this was still a problem in 2000 [6], and some technologists doubt that systems can ever be secure because of the existence of this attack [7]. Anderson et al. argue that the general risk of subversion is increasing [8].

Recently, in several mailing lists and blogs, a special technique to detect such attacks has been briefly described, which uses a second (diverse) "trusted" compiler and two compilation stages. This paper terms the technique "diverse double-compiling" (DDC). In DDC, if the final result is bit-for-bit identical to the original compiler binary, then the compiler source code accurately represents the binary. However, there is no peer-reviewed paper discussing DDC in detail, justifying its effectiveness, or discussing its ramifications. In addition, there is no public evidence that DDC has been tried, or a detailed description of how to perform it in practice. This paper resolves these problems.

This paper provides background and a description of the threat, followed by a description of DDC and a justification for its effectiveness. The next sections discuss how diversity can be used to increase trust in a second compiler and how to overcome practical challenges. The paper then presents a demonstration of DDC, and closes with ramifications.

## 2. Background

### 2.1. Inadequate solutions

Some simple approaches appear to solve the problem at first glance, yet have significant weaknesses:

1. Compiler binary files could be manually compared with their source code. This is impractical given compilers' large sizes, complexity, and rate of change.
2. Such comparison could be automated, but optimizing compilers make such comparisons difficult, compiler changes make keeping such tools up-to-date difficult, and the tool's complexity would be similar to a compiler's.
3. A second compiler could compile the source code, and then the binaries could be compared automatically to argue semantic equivalence. There is some work in determining the semantic equivalence of two different binaries [9], but this is very difficult.
4. Receivers could require that they only receive source code and then recompile everything themselves. This fails if the receiver's compiler is already malicious. An attacker could also insert the attack into the compiler's source; if the receiver accepts it (due to lack of diligence or conspiracy), the attacker could remove the evidence in a later version.
5. Programs can be written in interpreted languages. But eventually an interpreter must be implemented by machine code, so this simply moves the attack location.

### 2.2. Related work

Draper recommends screening out malicious compilers by writing a "paraphrase" compiler (possibly with a few dummy statements) or a different compiler binary, compiling once to remove the Trojan horse, and then compiling a second time to produce a Trojan-horse-free compiler [10]. This idea is expanded upon by McDermott [11], who notes that the alternative compiler could be a reduced-function compiler or one with large amounts of code unrelated to compilation. Lee's "approach #2" describes most of the basic process of diverse double-compilation (DDC), but implies that the results might not be bit-for-bit identical [12]. Luzar makes a similar point as Lee, describing how to rebuild a system from scratch using a different trusted compiler but not noting that the final result should be bit-for-bit identical if other factors are carefully controlled [13].

None of these works note that it is possible to produce a result that is bit-for-bit identical to the original compiler binary. This is an *essential* component of DDC; DDC gains significant advantages over other approaches because it is easy to determine if two files are exactly identical. These previous approaches require each defender to insert themselves into the compiler creation process (e.g., to recompile their compiler themselves before using it), which is often impractical. Resolving this by using a central trusted build agent simply moves the best point of attack. In contrast, DDC can be used for after-the-fact vetting by third parties, it does not require a fundamental change in the compiler delivery or installation process, it does not require that all compiler users receive or recompile compiler source code, and its evidence can be strengthened using multiple parties. Also, none of these papers demonstrate their technique.

Magdsick discusses using different versions of a compiler and different compiler platforms (CPU and operating system) to check binaries, but presumes that the compiler itself will simply be the same compiler (just a different version). He does note the value of recompiling "everything" to check it [14]. Anderson notes that cross-compilation does not help if the attack is in the compiler [15]. Mohring argues for the use of recompilation by gcc to check other components, presuming that the gcc binaries themselves in some environments would be pristine[16]. He makes no notice that all gcc implementations used might be malicious, or of the importance of diversity in compiler implementation. In his approach different compiler versions may be used, so outputs would be "similar" but not identical; this leaves the difficult problem of comparing binaries for "exact equivalence" unresolved.

Some effort has been made to develop proofs of correctness for compilers [17][18][19][20]. Goerigk argues that this problem requires that proofs of compilers go down to the resulting binary code [21] [22]. Such techniques are difficult to apply, even for simple languages.

There are a number of papers and articles about employing diversity to aid computer security, though they generally do not examine how to use diversity to counter Trojan horses inside compilers or the compilation environment. Geer et al. argue that a monoculture (an absence of diversity) in computing platforms is a serious security problem [23][24], but do not discuss employing compiler diversity to counter this particular attack. Forrest et al. argue that run-time diversity in general is beneficial for computer security [25]. In particular, their paper discusses techniques to vary final binaries by "randomized" transformations affecting compilation, loading, and/or execution. Their goal was to automatically change the binary (as seen at run-time) in some random ways sufficient to make it more difficult to attack. The paper provides a set of examples, including adding/deleting nonfunctional code, reordering code, and varying memory layout. They demonstrated the concept through a compiler that randomized the amount of memory allocated on a stack frame, and showed that the approach foiled a simple buffer overflow attack. This provides little defense against a subverted compiler, which can insert an attack before the countermeasures have a chance to thwart it.

Cowan et al. categorize "post hoc" techniques (adaptations to software after implementation to improve its security), based on what is adapted (the interface or the implementation) and on how it is adapted (either it is restricted or it is obscured). The paper does not specifically address the problem of malicious compilers [26].

Spinellis argues that "Thompson showed us that one cannot trust an application's security policy by examining its source code... The recent Xbox attack demonstrated that one cannot trust a platform's security policy if the applications running on it cannot be trusted" [27].

It is worth noting that the literature for change detection (such as [28]) and intrusion detection do not easily address this problem. Here the compiler is operating normally: it is expected to accept source code and generate object code.

Faigon's "Constrained Random Testing" detects compiler defects by creating many random test programs, compiling them with a compiler under test and a reference compiler, and detecting if running them produces different results [29]. This is extremely unlikely to find the Trojan horses considered here.

# 3. Analysis of threat

Thompson describes how to perform the attack, but there are some important characteristics of the attack that are not immediately obvious from his presentation. This section examines the threat in more detail and introduces terminology to describe it.

We'll begin by defining the threat. The threat considered in this paper is that an attacker may have modified one or more binaries (which computers run, but humans do not normally view) so that the compilation process inserts different code than would be expected from examining the compiler source code, sufficient so that recompilation of the compiler will cause the re-insertion of the malicious code. As a result, humans can examine the original source code without finding the attack, and they can recompile the compiler without removing the attack. For our purposes we'll call a subverted compiler a malicious compiler, and the entire attack the "trusting trust" attack.

Next, we'll examine what might motivate an attacker to actually perform such an attack, and the mechanisms an attacker uses that make this attack work (triggers, payloads, and non-discovery).

## 3.1. Attacker motivation

Understanding any potential threat involves determining the benefits to an attacker of an attack, and comparing them to the attacker's risks, costs, and difficulties. Although this "trusting trust" attack may seem exotic, its large benefits may outweigh its costs to some attackers.

The potential benefits are immense to a malicious attacker. A successful attacker can completely control all systems that are compiled by that binary and that binary's descendants, e.g., they can have backdoor passwords inserted for logins and gain unlimited privileges on entire classes of systems. Since detailed source code reviews will not find the attack, even defenders who have highly valuable resources and check all source code are vulnerable to this attack.

For a widely-used compiler, or one used to compile a widely-used program or operating system, this attack could result in global control. Control over banking systems, financial markets, militaries, or governments could be gained with a single attack. An attacker could possibly acquire limitless funds (by manipulating the entire financial system), acquire or change extremely sensitive information, or disable a nation's critical infrastructure on command.

An attacker can perform the attack against multiple compilers as well. Once control is gained over all systems that use one compiler, trust relationships and network interconnections could be exploited to ease attack against other compiler binaries. This would be especially true of a patient and careful attacker; once a compiler is subverted, it is likely to stay subverted for a long time, giving an attacker time to use it to launch further attacks.

An attacker (either an individual or an organization) who subverted a few of the most widely used compilers of the most widely-used operating systems could effectively control, directly or indirectly, almost every computer in existence.

The attack requires knowledge about compilers, effort to create the attack, and access (gained somehow) to the compiler binary, but all are achievable. Compiler construction techniques are standard Computer Science course material. The attack requires the insertion of relatively small amounts of code, so the attack can be developed by a single knowledgeable person in their spare time. Access rights to change the relevant compiler binaries might be harder to acquire, but there are clearly some who have such privileges already, and a determined attacker could acquire such privileges through a variety of means (including network attack, social engineering, physical attack, bribery, and betrayal).

The amount of power this attack offers is great, so it is easy to imagine a single person deciding to perform this attack for their own ends. An individual entrusted with compiler development might even succumb to the temptation if they believed they could not be caught, and the legion of virus writers shows that people are willing to write malicious code even without gaining the control this attack can provide.

Given such extraordinarily large benefits to an attacker, a highly resourced organization (such as a government) might decide to undertake it. Such an organization could supply hundreds of experts, working together full-time to deploy attacks over a period of decades. Defending against this scale of attack is beyond the ability of even many military organizations, and is far beyond the defensive abilities of the companies and non-profit organizations who develop and maintain popular compilers.

In short, this is an attack that can yield complete control over a vast number of systems, even those systems whose defenders perform independent source code analysis (e.g., those who have especially high-value assets), so this is worth defending against.

### 3.2. Triggers, payloads, and non-discovery

This attack depends on three things: triggers, payloads, and non-discovery. For purposes of this paper, a "trigger" is a condition determined by an attacker in which a malicious event is to occur (e.g., malicious code is inserted into a program). A "payload" is the code that actually performs the malicious event (e.g., the inserted malicious code and the code that causes its insertion). By "non-discovery," this paper means that victims cannot determine if a binary has been tampered with in this way; the lack of transparency in binary files makes this attack possible.

For this attack to be valuable, there must be at least two triggers: one to cause a malicious attack directly of value to the attacker (e.g., detecting compilation of a "login" program so that a Trojan horse can be inserted into it), and another to propagate attacks into future versions of the compiler.

If a trigger is activated when the attacker does not intend the trigger to be activated, the probability of detection increases. However, if a trigger is not activated when the attacker intends it to be activated, then that particular attack will be disabled. If all the attacks by the compiler against itself are disabled, then the attack will no longer propagate; once the compiler is recompiled, the attacks will disappear. Similarly, if a payload requires a situation that (through the process of change) disappears, then the payload will no longer be effective (and its failure may reveal the attack).

In this paper, "fragility" is the susceptibility of this attack to failure, i.e., that a trigger will activate when the attacker did not wish it to (risking a revelation of the attack), fail to trigger when the attacker would wish it to, or that the payload may fail to work as intended. Fragility is unfortunately less helpful to the defender than it might first appear. An attacker can counter fragility by simply incorporating many narrowly-defined triggers and payloads. Even if a change causes one trigger to fail, another trigger may still fire. By using multiple triggers and payloads, an attacker can attack multiple points in the compiler and attack different subsystems as final targets (e.g., the login system, the networking interface, and so on). Thus, there may be enough vulnerabilities in the resulting system to allow attackers to re-enter and re-insert new triggers and payloads into a malicious compiler. Even if a compiler misbehaves from malfunctioning malware, the results will often appear to be a mysterious compiler defect; if programmers "code around" the problem, the attack will stay undetected.

Since attackers do not want their malicious code to be discovered, they may limit the number of triggers/payloads they insert and the number of attacked compilers. In particular, attackers may tend to attack only "important" compilers (e.g., compilers that are widely-used or used for high-asset projects), since each compiler they attack (initially or to add new triggers and payloads) increases the risk of discovery. However, since these attacks can allow an attacker to deeply penetrate systems generated with the compiler, malicious compilers make it easier for an attacker to re-enter a previously penetrated development environment to refresh a binary with new triggers and payloads. Thus, once a compiler has been subverted, it may be difficult to undo the damage without a process for ensuring that there are no attacks left.

The text above might give the impression that only the compiler binary itself can influence results (or how they are run), yet this is obviously not true. Assemblers and loaders are excellent places to place a trigger (the popular gcc compiler actually generates assembly language as text and then invokes an assembler). An attacker could place the trigger mechanism in the compiler's supporting infrastructure such as the operating system kernel, libraries, or privileged programs. In many cases writing triggers is more difficult for such components, but in some cases (such as I/O libraries) this is fairly easy to do.

## 4. Diverse double-compiling (DDC)

The idea of diverse double-compiling (DDC) was first created and posted by Henry Spencer in 1998 [30], inspired by McKeeman et al's exercise for detecting compiler defects [31][32]. Since this time, this idea has been posted in several places, all with very short descriptions [16][33][34].

To perform DDC, recompile a compiler's source code twice: once with a second "trusted" compiler, and again using the result of the first compilation. Then, check if the final result exactly matches the original compiler binary; if it does, then there is no Trojan horse in the binary (given some assumptions to be discussed later). This technique uses the second (trusted) compiler as a check on the first. Thompson's attack assumes that there is only one compiler available; adding a second compiler invalidates this assumption. The trusted compiler and its environment may be malicious, as long as that does not impact their result in this case, and they may be very slow.

Figure 1 illustrates the process of DDC in more detail, along with a self-regeneration check. This figure shows binary file(s) for an untrusted compiler A, binary file(s) for a trusted compiler T, and source code $s_A$ that is claimed to be the source code of compiler A. The result of compiling source SC using compiler X is notated as c(SC,X). The shaded boxes show a compilation step; in this notation, a compilation uses a compiler (input from the top), source code (input from the left), and other data (input from the right), all to produce a binary (output exiting down). File comparisons are shown as labeled dashed lines.

Before performing DDC, we should first do a regeneration check. This check acts like the control of an experiment; it detects when a compiler cannot regenerate itself. Simply take source code $s_A$ and compile it with compiler A, producing binary file c($s_A$,A), that is, source $s_A$ compiled by compiler A. We then do a bit-for-bit comparison (Compare1) to see if c($s_A$,A) is the same as A. If c($s_A$,A) is the same as A, then the compiler can regenerate (reproduce) itself. This does not prove the absence of malice, however.

We then perform DDC. We start by using trusted compiler T to compile $s_A$ to produce c($s_A$,T), that is, source $s_A$ compiled by T. We then use c($s_A$,T) to compile $s_A$ again, producing the binary c($s_A$,c($s_A$,T)). The final result is then compared (in a trusted environment) to the original A and c($s_A$,T); if c($s_A$,c($s_A$,T)), A, and c($s_A$,T) are identical, then we can say that $s_A$ accurately reflects A (we'll see why this is so in the next section). These two compilation steps will be called stage 1 and stage 2, and are the origin of its name: we compile twice, the first time using a different (diverse) compiler. All three compilations (self-regeneration check, stage 1, and stage 2) could be performed on the same or on different environments.

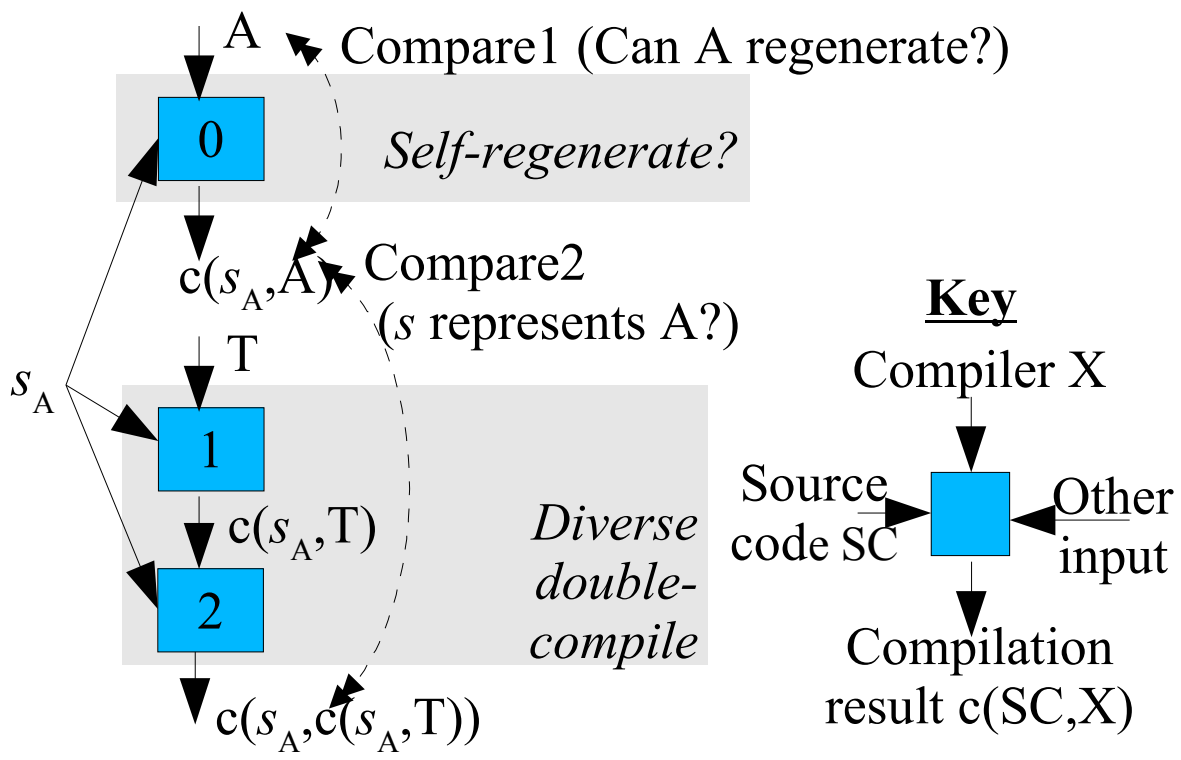

**Fig. 1. Diverse double-compiling with self-regeneration check**

## 5. Justification

To justify this technique, we must first state some assumptions:

1. We must have a trusted compilation process T, comparer, and environment(s) used in DDC, and trusted way to acquire A and $s_A$. "Trusted" here means we have reason to believe it does not have triggers and payloads that affect those actions identically as possible triggers and payloads in the compiler under test. They may have triggers and payloads, but they do not matter if they do not affect the result, and defects are likely to be detected. Justifying this assumption is discussed in section 6.
2. T must have the same semantics for the same constructs as A does, for the set of constructs used in source code $s_A$. Obviously, a Java™ compiler cannot be used directly as T if $s_A$ is written in the C language! But if $s_A$ uses any nonstandard language extensions, or depends on a construct not defined by a language specification, then T must implement them in the way expected by $s_A$. If a different environment is used, additional challenges may arise (e.g., byte ordering problems) unless $s_A$ was designed to be portable. Any defect in T can also cause problems, though defects will be detected by the process unless they do not affect $s_A$ or A has exactly the same defects with the same semantic results. Only the semantics need to be identical; T may be very slow, run on a different processor or virtual machine, and produce code for a different processor or virtual machine.

3. The information (such as option flags) that affects the output of compilation must be semantically identical when generating c($s_A$,A) and c($s_A$,c($s_A$,T)). If the stage 2 environment is different from the self-regeneration stage's, minor porting may be needed. Any input such as command line parameters (including option flags that change the results), important environment variables, libraries used as data, and so on that affect the outcome must be controlled.

4. The compiler defined by $s_A$ should be deterministic given only its inputs, and not use or write undefined values. Given the same source code and other inputs, it should produce exactly the same outputs. If the compilation is non-deterministic, in some cases it could be handled by running the process multiple times, but in practice it is easier to control enough inputs to make the compiler deterministic. Non-determinism hides other problems, in any case, and makes finding flaws much more difficult; uncontrolled non-determinism in a compiler should be treated as a defect. Although undefined values may be deterministic in a particular environment, if the environment changes the undefined values may also change, with the same result. It may be possible to work around this by carefully setting undefined values to a defined value, but it is better to fix the compiler to not do this in the first place. If timestamps are embedded in the output, the time should be controlled so that they will be identical in both outputs or some alternative approach (discussed later) must be used.

The "self-regeneration" step is important; until we can reproduce the binary of the compiler A using itself, we cannot hope to reproduce it with a more complicated process. This step is likely, for example, to detect non-determinism in A. If A does not regenerate itself, then we must first determine how to repeatably generate A.

We can now make the following assertions, if the preceding assumptions are true:

1. Stage 1's result, c($s_A$,T), will be functionally the same as A if $s_A$ represents A. Stage 1 simply compiles $s_A$ using T to produce program c($s_A$,T). c($s_A$,T) will normally have a different representation than A, since it was compiled using the different compiler T. Indeed, T may generate code for a completely different processor. But if source $s_A$ truly represents the source of compiler A, and the other assumptions are true, then c($s_A$,T) will be functionally the same as A. E.g., if $s_A$ is an x86 compiler, compiler A an x86 binary, and T generates 68K code, then c($s_A$,T) would run on a 68K —but since $s_A$ is for an x86 compiler, running c($s_A$,T) would generate x86 code.

2. Even if A is malicious, it cannot affect the result c($s_A$,c($s_A$,T)). During DDC, program A (which is potentially malicious) is never used at all. Instead, during DDC we only use a trusted compilation process T, code generated by T, and other programs in environments which we trust do not trigger on compilation of $s_A$. Thus, even if A is malicious, it cannot affect the outcome.

3. Stage 2's result, c($s_A$,c($s_A$,T)), will be identical to c($s_A$,A) and A iff $s_A$ accurately represents A. Since c($s_A$,T) is supposed to be functionally the same as A, we can execute c($s_A$,T) to compile the original source code $s_A$, producing yet another new binary c($s_A$,c($s_A$,T)). But since this new binary was compiled with a program that is supposed to be functionally identical to A, and all other compilation inputs that affected compilation results were kept the same, then its output should be the same as A... and since the input is $s_A$, the output should be the same as A and c($s_A$,A). Continuing the assertion 1 example, c($s_A$,T) will generate x86 code the same way A is supposed to, so if it is given $s_A$, it should produce A. If c($s_A$,c($s_A$,T)) is different from A, then at least one assumption listed above is false or A has been changed in a way not visible in $s_A$ (e.g., by having malicious content).

Note key limitations of this technique:

1. It only shows that the source and binary files correspond, i.e., that there is "nothing hidden." The source code may have Trojan horses and errors, in which case the binary file will too. However, if the source and binary correspond, the source code can be analyzed in the usual ways to find such problems.

2. It only shows that a particular binary and source code match. There may be other binaries that contain Trojan horse(s) not represented by the source, but they will be different in some way.

## 6. Methods to increase diversity

DDC requires a trusted compiler T and trusted environment(s) where there is a high degree of confidence that any triggers against $s_A$ that may be in compiler A will not also be present. Trust can be gained in a variety of ways; one way is to perform a complete formal proof of compiler T's implementation and of the environments used in DDC, along with evidence that what actually runs is what was proved.

A simpler method to gain a great amount of trust is through diversity, and there are many ways we can gain diversity to increase the claim's strength. These include diversity in compiler implementation, in time, in environment, and in input source code.

### 6.1. Diversity in compiler implementation

Ideally, compiler T's binary should be for a completely different implementation than of compiler A. Compiler T's binary could include triggers and payloads for other compilers (such as compiler A), but this is much less likely, since an attacker would then have to subvert the development process of multiple compiler binaries to do so.

Ideally, compiler T has never been compiled by any version of compiler A, even in T's initial bootstrap. This is because compiler A could insert into the binary code some routines to check for any processing of compiler A (itself), so that it can later "re-infect" itself. This kind of attack is difficult to do, however, especially since bootstrapping is usually done very early in a compiler's development and an attacker may not even be aware of the compiler T's development at that time. One of the most obvious locations where this might be practical might be in the I/O routines. However, I/O routines are more likely to be viewed at the assembly level (e.g., to do performance analysis), so an attacker risks discovery if they subvert I/O routines.

### 6.2. Diversity in time

If compiler T and the DDC environment were developed long before the compiler A, and they do not share a common implementation heritage, it is improbable that compiler T or its environment would include relevant triggers for a not-yet-implemented compiler (Magdsick makes a similar point [14]). It is possible that an attacker could arrange to include triggers in compiler A's source code once compiler A is developed, but this is extremely difficult to do, and is even more difficult to maintain over time as compilers change.

Using a newer compiler binary to check an older compiler gains less confidence; it is easier for a recently-released compiler binary to include triggers and payloads for many older compilers, including completely different compilers. Still, this requires the subversion of multiple different compilers' binaries, so even this case can increase confidence.

Diversity achieved via earlier development can only provide significant confidence if it can be clearly verified that compiler T and/or the DDC environments are truly the ones that existed at the earlier time. In particular, old versions should not be simply acquired over the Internet without independent verification, because a resourceful attacker could tamper with those copies. Instead, protected copies of the original media should be preferred to reduce the risk of tampering. Other copies can be used to verify that the data used is correct. Cryptographic hashes can be used to verify the media; multiple hash algorithms should be used, in case a hash algorithm is broken.

An older binary version of compiler A can be used as compiler T, if there is reason to believe that the old version is not malicious or that any Trojan horse in the old version of A will not be triggered by $s_A$. Note that this is a weaker test; the common ancestor could have been subverted. This technique gives greater confidence if the changes in the compiler have been so significant that the newer version is in essence a different compiler, but it would be best if compiler T were truly a separate implementation.

### 6.3. Diversity in environment

Different environments could be used. Compiler T could generate code for a different environment; T and/or c($s_A$,T) could run on a different environment. The term "environment" here means the entire infrastructure supporting the compiler including the CPU architecture, operating system, supporting libraries, and so on. It should not be running any other processes (which might try to use kernel vulnerabilities to detect a compilation and subvert it). Using a completely different environment counters Trojan horses whose triggers and payloads are actually in the binaries of the environment, as well as countering triggers and payloads that only work on a specific operating system or CPU architecture.

These benefits could be partly achieved through emulation of a different system. There is always the risk that the emulation system or underlying environment could be subverted specifically to give misleading results, but attackers will find this difficult to achieve, particularly if the emulation system is developed specifically for this test (an attacker might have to develop the attack before the system was built!).

### 6.4. Diversity in source code input

Another way to add diversity would be to use mutations of compiler A's source code as the input to the first stage of DDC [10][11]. Compiler T is then a source code transform, a compiler (possibly the original compiler), and possibly a postprocessing step.

Semantic-preserving mutations change the source code without changing its semantics. This could include actions such as renaming items (such as variables, functions, and/or filenames), reordering statements where the order is irrelevant, regrouping statements, intentionally performing unnecessary operations that will not produce an output, changing to different algorithms that produce sufficiently similar results, and changing compiler opcode values for internal data structures. Even trivial changes, such as changing whitespace, increases diversity (these trivial changes can still be enough to counter triggers if those triggers depend on them). Forrest discusses several methods for introducing diversity [25]. McDermott notes that even changed semantics are helpful, e.g., performing excess tasks whose results are ignored [11].

By inserting such mutations, it is less likely that triggers designed to attack compiler A will activate in the compiler used inside T, and if they do, the payloads in compiler T are less likely to be effective. These mutations could be implemented by automated tools, or even manually. Since it is part of T, trust is given to the mutator (be it manual or automated). If the mutator has an unintentional defect, the result will be simply that a difference will be identified; tracking backwards to explain the difference will identify the defect, so defects in the mutator are not as serious.

## 7. Practical challenges

There are many practical challenges to implementing this technique, but they can generally be overcome.

Uncontrolled nondeterminism or using uninitialized data may cause a compiler to generate different answers for the same source input. It may be easiest to modify the compiler so that it can be made to be deterministic (e.g., add an option to set a random number seed) and to never use uninitialized data. Differences that do not affect the outcome are fine, e.g., heap memory allocations during compilation often allocate different memory addresses, but this is only a problem if the compiler output changes depending on those addresses' specific values. Roskind reports that variance in heap address locations affected the output of at least some versions of the Javasoft javac compiler. He also stated that he felt that this was a bug, noting that this behavior made port validation extremely difficult [35].

It may be difficult to compile $s_A$ using existing trusted compilers. Thankfully, there are many possible solutions if $s_A$ cannot be compiled by a given trusted compiler. An existing trusted compiler could be modified (e.g., to add extensions) so it can compile $s_A$. Another alternative is to create a trusted preprocess step that is applied to $s_A$, possibly done by hand; as a result T would be defined as being the preprocess step plus the trusted compiler. Trusted compiler T could be created by using an existing trusted compiler (but one that cannot compile $s_A$ directly) to compile another existing trusted compiler that can compile $s_A$, i.e., the first trusted compiler is used to bootstrap another compiler. It is possible to write a new trusted compiler from scratch; since performance is irrelevant and it only needs to be able to compile one program, this may not be difficult. An old version of A could be used as T, but that is far less diverse so the results are far less convincing, and risks "pop-up" attacks.

A "pop-up" attack, as defined in this paper, is where an attacker includes a self-perpetuating attack in only some versions of the source code (where the attack "pops up"), with the idea that defenders may not examine the source code of those particular versions in detail. Imagine that T is used to determine that an old version of compiler A (call it A1) corresponds to its source $s_{A1}$. Now imagine that an attacker cannot modify binaries directly (e.g., because they are regenerated by a suspicious user), but that the attacker can modify the source code of the compiler (e.g., by breaking into its repository). The attacker could sneak malevolent self-perpetuating code into $s_{A2}$ (which is used to generate A2), and then remove that malevolent code from $s_{A3}$. If A2 is used to generate A3, then A3 may be malicious, even though examining $s_{A3}$ will not reveal an attack. Examination of every change in the source code at each stage can prevent this, but this must be thorough; examining only the source's beginning and end-state will miss the attack. It is safer to re-run DDC on every release; if that is impractical, at least do it periodically to reduce the attack window.

Compilers may have multiple subcomponents. It may be necessary to break $s_A$ into subcomponents and handle them separately, possibly in a certain order to address dependencies. Section 8 demonstrates this.

Inexact comparisons may be needed. The comparisons (Compare1 and 2) need not require an identical result as long as it can be shown that the differences that do not cause a change in behavior. This might occur if, for example, outputs included embedded compilation timestamps. However, showing that differences in files do not cause differences in the functionality, in the presence of an adversary, is extremely difficult. An alternative that can work in some cases is to run additional self-generation stages until a stable result occurs. Another approach is to first work to make the results identical, and then show that the steps leading from that trusted point do not introduce an attack.

The environment of A may be untrusted. As noted earlier, an attacker could place the trigger mechanism in the compiler's supporting infrastructure such as the operating system kernel, libraries, or privileged programs. Triggers would be especially easy to place in assemblers, linkers, and loaders. But even unprivileged programs might be enough to subvert compilations; an attacker could create a program that exploited unknown kernel vulnerabilities. The DDC technique can be used to cover these cases as well. Simply redefine A as the set of all components to be checked; this could even be the set of all software that runs on that machine (including all software run at boot time). This means that the source code for all this software to be checked is $s_A$. Consider obtaining A and $s_A$ from some read-only medium (e.g., CD-ROM or inactive hard drive); do not trust A to produce itself (e.g., by copying A's files using A)! Then, using DDC on a different (trusted) environment, rebuild A using $s_A$; in the limit this would regenerate all of the operating system (including boot software), application programs, and so on. Files that are directly reviewed by humans (e.g., interpreted non-binaries) can be "compiled" to themselves. If DDC can regenerate the original A, then the entire set of components included in A are represented by the entire set of source code in $s_A$. If A or its environment might have code that shrouds $s_A$, always use a trusted system to view/print $s_A$ when examining $s_A$.

A resourceful attacker might attack the system performing DDC (e.g., over a network) to subvert its results. DDC should be done on isolated system(s). Ideally, the systems used to implement DDC should be rebuilt from trustworthy media, not connected to external networks at all, and not run any programs other than those necessary for the test.

Few will want to do DDC themselves. This technique might be difficult to do the first time for some compilers, and in any case there is no need for everyone to perform this check. Organization(s) trusted by many others (such as government agencies or trusted organizations sponsored by them) could perform these techniques on a variety of compilers/environments, as they are released, and report the cryptographic hash values of the binaries and their corresponding source code. The source code would not need to be released to the world, so this technique could be applied to proprietary software. This would allow others to quickly check if the binaries they received were, in fact, what their software developers intended to send. If someone did not trust those organizations, they could ask for another organization they did trust to do this (including themselves, if they can get the source code). Organizations that do checks like this have elsewhere been termed "trusted build agents" [16].

# 8. Demonstration using tcc

There is no public evidence that this technique has been used. One 2004 gcc mailing list posting stated, "I'm not aware of any ongoing effort," [36]; another responded, "I guess we all sorta hope someone else is doing it." [37]. This section describes its first demonstration.

A public demonstration requires a compiler whose source code is publicly available. Other ideal traits for the initial test case included being relatively small and self-contained, running quickly (so that test runs would be rapid), having an open source software license (so the experiment could be repeated and changes could be publicly redistributed [38]), and being easily compiled by another compiler. The compiler needed to be relatively defect-free, since defects would interfere with these tests. The Tiny C Compiler, abbreviated as TinyCC or tcc, was chosen as it appeared to meet these criteria.

The compiler tcc was developed by Fabrice Bellard and is available from its website at http://www.tinycc.org/. This project began as the Obfuscated Tiny C Compiler (OTCC), a very small C compiler Bellard wrote to win the International Obfuscated C Code Contest (IOCCC) in 2002. He then expanded this small compiler so that it now supports all of ANSI C, most of the newer ISO C99 standard, and many GNU C extensions including inline assembly. The compiler tcc appeared to meet the requirements given above. In addition, tcc had been used to create "tccboot," a Linux distribution that first booted the compiler and then recompiled the entire kernel as part of its boot process. This capability to compile almost all code at boot time could be very useful for future related work, and suggested that the compiler was relatively defect-free.

The following sections describe the test configuration, the DDC process, problems with casting 8-bit values and long double constants, and final results.

## 8.1. Test configuration

All tests ran on an x86 system running Red Hat Fedora Core 3. This included Linux kernel version 2.6.11-1.14_FC3 and gcc version 3.4.3-22.fc3. gcc was both the bootstrap compiler and the trusted compiler for this test; tcc was the simulated potentially malicious compiler.

First, a traditional chain of recompilations was performed using tcc versions 0.9.20, 0.9.21, and 0.9.22. After bootstrapping, a compiler would be updated and used to compile itself. Their gzip compressed tar files have the following SHA-1 values (provided so others can repeat this experiment):

```
6db41cbfc90415b94f2e53c1a1e5db0ef8105eb8  0.9.20
19ef0fb67bbe57867a590d07126694547b27ef41  0.9.21
84100525696af2252e7f0073fd6a9fcc6b2de266  0.9.22
```

As is usual, any such sequence must start with some sort of bootstrap of the compiler. gcc was used to bootstrap tcc-0.9.20, causing a minor challenge: gcc 3.4.3 would not compile tcc-0.9.20 directly because gcc 3.4.3 added additional checks not present in older versions of gcc. In tcc-0.9.20, some functions are declared like this, using a gcc extension to C:

```
void *__bound_ptr_add(void *p, int offset)
__attribute__((regparm(2)));
```

but the definitions of those functions in tcc's source code omit the __attribute__((regparm(...))). gcc 3.4.3 perceives this as inconsistent and will not accept it. Since this is only used by the initial bootstrap compiler, we can claim that the bootstrap compiler has two steps: a preprocessor that removes these regparm statements, and the regular gcc compiler. The regparm text is only an optimization with no semantic change, so this does not affect our result.

This process created a tcc version 0.9.22 binary file which we have good reasons to believe does not have any hidden code in the binary, so it can be used as a test case. Now imagine an end-user with only this binary and the source code for tcc version 0.9.22. This user has no way to ensure that the compiler has not been tampered with (if it has been tampered with, then its binary will be different, but this hypothetical end-user has no "pristine" file to compare against). Would DDC correctly produce the same result?

## 8.2. Diverse double-compiling tcc

Real compilers are often divided into multiple pieces. Compiler tcc as used here has two parts: the main compiler (file tcc) and the compiler run-time library (file libtcc1.a; tcc sometimes copies portions of this into its results). For purposes of this demonstration, these were the only components being checked; everything else was assumed to be trustworthy for this simple test (this assumption could be removed with more effort). The binary file tcc is generated from the source file tcc.c and other files; this set is notated $s_{tcc}$. Note: the tcc package also includes a file called tcclib, which is not the same as libtcc1.

Figure 2 shows the process used to perform DDC with compiler tcc. First, a self-regeneration test was performed to make sure we could regenerate files tcc and libtcc1; this was successful. Then DDC was performed. Notice that stages one and two, which are notionally one compilation each, are actually two compilations each when applied to compiler tcc because we must handle two components in each stage (in particular, we must create the recompiled run-time before running a program that uses it).

One challenge is that the run-time code is used as an archive format (.a format), and this format includes a compilation timestamp of each component. These timestamps will, of course, be different from any originals unless special efforts are made. Happily, the runtime code is first compiled into an ELF .o format (which does not include these timestamps), and then transformed into an archive format using a trusted program (ar). So, for testing purposes, the libtcc1.o files were compared and not the libtcc1.a files.

Unfortunately, when this process was first tried, the DDC result did not match the result from the chain of updates, even when only using formats that did not include compilation timestamps. After much effort this was tracked to two problems: a compiler defect in sign-extending values cast to 8-bit values, and uninitialized data used while storing long double constants. Each of these issues is discussed next, followed by the results after resolving them.

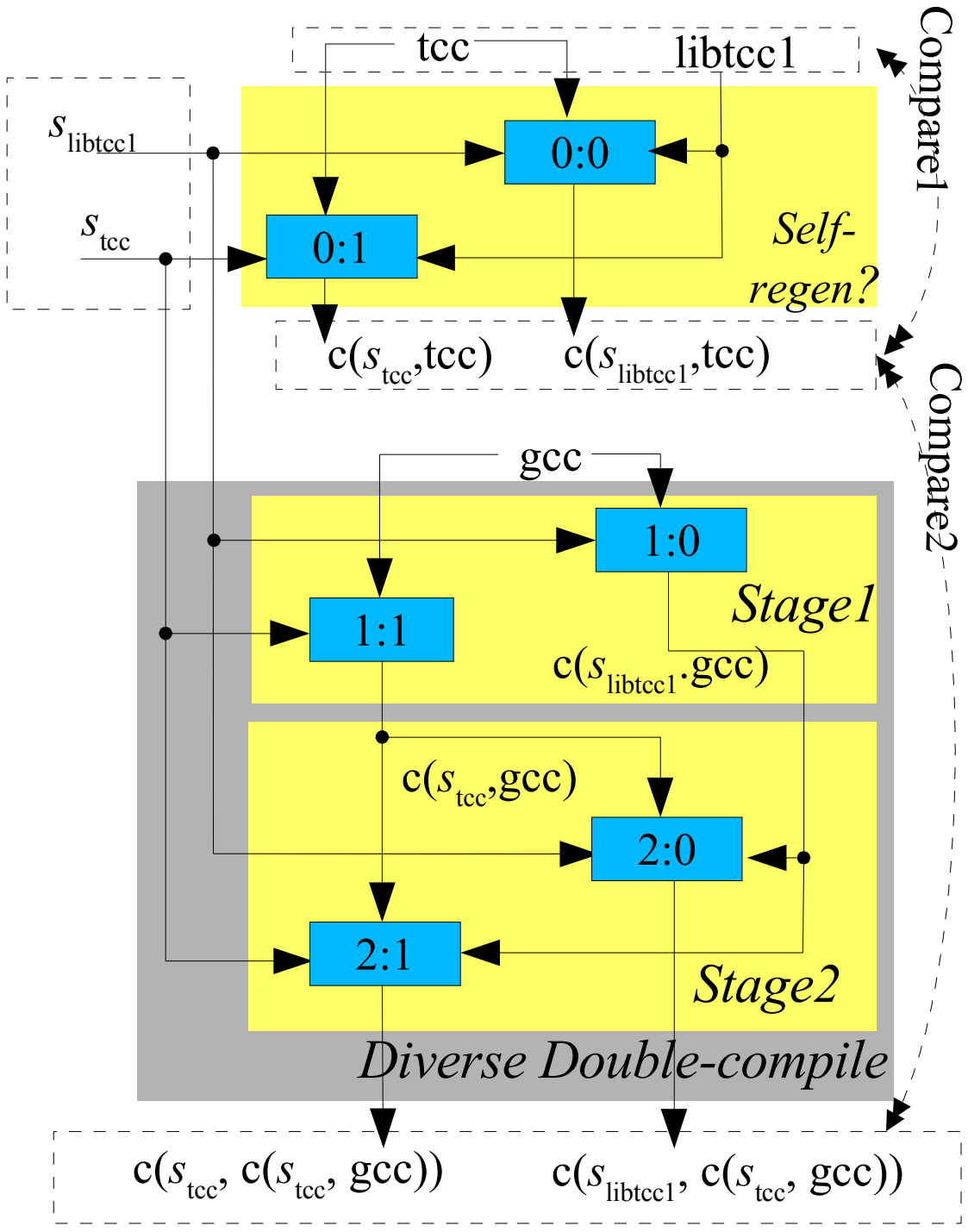

**Fig. 2. Diverse double-compiling with self-regeneration check, using tcc**

## 8.3. Defect in sign-extending cast 8-bit values

A subtle defect in tcc caused serious problems. The defect occurs if a 32-bit unsigned value is cast to a signed 8-bit value, and then that result is compared to a 32-bit unsigned value without first storing the result in a variable (which should sign-extend the 8-bit value). Here is a brief description of why this construct is used, why it is a defect, and the impact of this defect.

The x86 processor machine instructions can store 4 byte constants as 4 bytes, but since many such constants are in the range -128..127, constants in this range can also be stored in a shorter 1-byte format (by specifying a specific ModR/M value in the machine instruction). Where possible, tcc tries to use the shorter form, using statements like this to detect them (where e.v is of type uint32, an unsigned 32-bit value):

```
if (op->e.v == (int8_t)op->e.v && !op->e.sym) {
```

Unfortunately, the value cast to (int8_t) is not sign-extended by tcc version 0.9.22 when compared to an unsigned 32-bit integer. Version 0.9.22 does drop the upper 24 bits on the first cast to the 8-bit signed integer, but it fails to sign-extend the remaining 8-bit signed value unless the 8-bit value is first stored in a variable. This is a defect, at least because tcc's source code depends on a drop with sign-extension and tcc is supposed to be self-hosting. It is even more obvious that this is a defect because using a temporary variable to store the intermediate result *does* enable sign-extension. Besides, this is documented as a known defect in tcc 0.9.22's own TODO documentation, though this was only discovered after laboriously tracking down the problem. According to Kernighan [39] section A6.2 and the ISO/IEC C99 standard section 6.3.1.3 [40], converting to a smaller signed type is implementation-defined, but conversion of that to a larger unsigned value should sign-extend. Note that gcc does do the drop and sign-extension (as tcc's author expects).

This defect results in incorrect code being generated by tcc 0.9.22 if it is given values in the range 0x80..0xff in this construct. But when compiling itself, tcc is lucky and merely generates slightly longer code than necessary in certain cases. Thus, a gcc-compiled tcc generates code of this form (where 3-byte codes are used) when compiling some inline assembly in the tcc runtime library libtcc1:

```
1b5: 2b 4d dc  sub 0xffffffdc(%ebp),%ecx
1b8: 1b 45 d8  sbb 0xffffffd8(%ebp),%eax
```

But a tcc-compiled tcc incorrectly chooses the "long" form of the same instructions (which have the same effect—note the identical disassembly):

```
1b5: 2b 8d dc ff ff ff  sub 0xffffffdc(%ebp),%ecx
1bb: 1b 85 d8 ff ff ff  sbb 0xffffffd8(%ebp),%eax
```

One of the key assumptions in DDC is that the two compilers agree on the semantics of the language being compiled. This tcc defect violates this assumption, causing the files to unexpectedly differ. To resolve this, tcc was modified slightly so it would store such intermediate values in a temporary variable, avoiding the defect; a better long-term solution would be to fix the defect.

This example shows that DDC can be a good test for unintentional compiler defects—small defects that might not be noticed elsewhere may immediately surface!

### 8.4. Long double constant problem

Another problem resulted from how tcc outputs long double constants. The tcc outputs floating point constants in the "data" section, but when tcc compiles itself, the tcc.c line:

```
if (f2 == 0.0) {
```

outputs inconsistent data section values to represent 0.0. The tcc compiled by gcc stores 11 0x00 bytes followed by 0xc9, while tcc compiled by itself generates 12 0x00 bytes. Because f2 has type "long double," tcc eventually stores this 0.0 in memory as a long double value. The problem is that tcc's "long double" uses only 10 bytes, but it is stored in 12 bytes, and tcc's source code does not initialize the extra 2 bytes. The two excess "junk" bytes end up depending on the underlying environment, causing variations in the output [41]. In normal operation these bytes are ignored and thus cause no problems.

To resolve this, the value "0.0" was replaced with the expression (f1-f1), since f1 is a long double variable known to have a finite value there (e.g., it is not a NaN). This is semantically the same and eliminated the problem. A better long-term solution for tcc would be to always set these "excess" values to constants (such as 0x00).

### 8.5. Final results with tcc demonstration

After patching tcc 0.9.22 as described above, and running it through the processes described above, exactly the same files were produced through the chain of updates and through DDC. This is shown by these SHA-1 hash values for the compiler and its runtime library, which were identical for both processes:

```
c1ec831ae153bf33bff3df3c248b12938960a5b6 tcc
794841efe4aad6e25f6dee89d4b2d0224c22389b libtcc1.o
```

But can we say anything about unpatched tcc 0.9.22? We can, once we realize that we can (for test purposes) pretend that the patched version came first, and that we then applied changes to create the unpatched version. Since we have shown that the patched version's source accurately represents the binary identified above, we only need to examine the effects of a reversed change that "creates" the unpatched version. Visual inspection of the reversed change quickly shows that it has no malicious triggers and payloads. Thus, we can add one more chain from the trusted compiler to a "new" version of the compiler that is the untouched tcc-0.9.22. Because of the changes in semantics and the flow of data, to get a stable result we end up needing to recompile several times. In the end, the following SHA-1 hash values are the correct binaries for tcc-0.9.22 on an x86 in this environment when tcc is self-compiled a sufficient number of times to become "stable":

```
d530cee305fdc7aed8edf7903d80a33b6b3ee1db tcc
42c1a134e11655a3c1ca9846abc70b9c82013590 libtcc1.o
```

## 9. Ramifications

This paper has summarized and demonstrated how to detect Thompson's "Trusting Trust" attack, using diverse double-compiling (DDC). This technique has many strengths: it can be completely automated, applied to any compiled language (including common languages like C), and does not require the use of complex mathematical proof techniques. Second-source compilers and environments are desirable for other reasons, so they are often already available, and if not they are also relatively easy to create (since high performance is unnecessary). Some unintentional defects in either compiler are also detected by the technique. The technique can be easily expanded to cover all of the software running on a system (including the operating system kernel, bootstrap software, libraries, microcode, and so on) as long as its source code is available.

Passing this test when the trusted compiler and environment is not proven is not a mathematical proof, but more like a legal one. The test can be made as strong as you wish, by decreasing the likelihood (e.g., through diversity) that the DDC process (including trusted compiler T and the environments) also have the malicious code. Multiple diverse DDC tests can strengthen the evidence even further. A defender can easily make it extremely unlikely that an attacker could subvert the DDC technique.

Note that this technique only shows that the source code corresponds with a given compiler's binary, i.e., that nothing is hidden. The binary may have errors or malevolent code; this technique simply ensures that these errors and malevolent code *can* be found by examining the source code. Passing this test makes source code analysis more meaningful.

As with any approach, this technique has limitations. The source code for the compiler being tested must be available to the tester, and the results are more useful to those who have access to the source code of what was tested (the compiler and/or the environment under test). Since the technique requires two compilers to agree on semantics, this is easier to do for popular languages where there is a public language specification and where no patents inhibit the creation of a second implementation. The technique is far simpler if the compiler being tested was designed to be portable and avoids using nonstandard extensions. It can be applied to microcode and hardware specification data as well, but applying it directly to hardware (like CPUs) requires an "equality" operation for hardware, which is more challenging.

This technique does have potential policy implications. To protect themselves and their citizenry, governments could enact policies requiring that they receive all of the source code (including build instructions) necessary to rebuild a compiler and its entire environment, and for it to be sufficiently portable so it can be built with an alternative trusted compiler and environment. Multiple compilers are easier to acquire for standardized languages, so governments could insist on the use of standard languages, specified in legally unencumbered public standards and implemented by multiple vendors, to implement compilers and critical infrastructure. Organizations (such as governments) could establish groups to do this testing and report the cryptographic hashes of corresponding binaries and source.

Future potential work includes examining a larger and more popular compiler (such as gcc), including an entire operating system as the "compiler A" under test, relaxing the requirement for exact equivalence, and demonstrating DDC with a more diverse environment (e.g., by using a much older operating system and different CPU architecture).

All URLs retrieved as of Sep. 18, 2005. Manuscript submitted May 27, 2005, and revised Sep. 22, 2005. This work was supported by IDA under its Central Research Program and written using OpenOffice.org. It is dedicated to the memory of Dennis W. Fife.